\documentclass[english,preprint]{aastex}
\usepackage[T1]{fontenc}
\usepackage[latin9]{inputenc}
\usepackage{babel}

\begin{document}

\title{Observational Determination of the Turbulent Ambipolar Diffusion
Scale and Magnetic Field Strength in Molecular Clouds}

\author{Talayeh Hezareh\altaffilmark{1}, \email{thezareh@uwo.ca} Martin
Houde\altaffilmark{1}, Carolyn M$^{\mathrm{c}}$Coey \altaffilmark{1,2},
and Hua-bai Li\altaffilmark{3}}

\altaffiltext{1}{The Department of Physics and Astronomy, The University of Western Ontario, London, Ontario, Canada N6A 3K7}
\altaffiltext{2}{Department of Physics and Astronomy, University of Waterloo, 200 University Avenue W., Ontario, Canada N2L 3G1}
\altaffiltext{3}{Max-Planck-Institut für Astronomie, Königstuhl 17, D-69117 Heidelberg, Germany}
\begin{abstract}
We study the correlation of the velocity dispersion of the coexisting
molecules $\mathrm{H^{13}CN}$ and $\mathrm{H^{13}CO^{+}}$ and the
turbulent energy dissipation scale in the DR21(OH) star-forming region.
The down-shift of the $\mathrm{H^{13}CO^{+}}$ spectrum relative to
$\mathrm{H^{13}CN}$ is consistent with the presence of ambipolar
diffusion at dissipation length scales that helps the process of turbulent
energy dissipation, but at a different cut-off for ions compared to
the neutrals. We use our observational data to calculate a turbulent
ambipolar diffusion length scale $L^{'}\simeq17$ mpc and a strength
of $B_{\mathrm{pos}}\simeq1.7$ mG for the plane of the sky component
of the magnetic field in DR21(OH). 
\end{abstract}


\section{Introduction}

Interstellar magnetic fields and turbulence have been the most debated
mechanisms of support against gravitational collapse in molecular
clouds. Recent studies have evolved towards acknowledging both mechanisms
by studying turbulence in a magnetized gas (e.g., \citealt{Shebalin,goldreich,ostriker,biskamp,basu,Falceta}).
It is, however, a challenge to combine the existing models for turbulence
with observational data for the measurement of the interstellar magnetic
fields to fully understand the dynamics in molecular clouds. There
is yet no closed theory for magnetized turbulence due to its complicated
and intermittent nature. On the other hand, interstellar magnetic
fields are weak ($\sim$ few hundred $\mu$G to a few mG) and difficult
to measure with the very few observational techniques currently at
hand (e.g., \citealt{crutcher2,Houde,Falgarone,Hildebrand,houde2009}).
It is therefore essential to innovate with more observing techniques
to provide reliable data and test existing models.

Turbulence is basically a dissipative process. In an incompressible
turbulent fluid, energy is transferred from large eddies to smaller
ones, until it is dissipated by viscosity at small enough scales \citep{Kolmogorov}.
The dissipation of turbulent energy has important effects on the ISM,
such as initiating collapse in supercritical cores \citep{basu},
modifying the chemical evolution of the gas molecules \citep{xie},
and causing the decoupling of gas and grains in a cloud \citep{fal.puget}. 

\citet{li&houde} established a new observational technique, expanding
on the original work of Houde et al. \citeyearpar{houde2000a,houde2000b}
on the effect of magnetic fields on the spectral line widths of ion
and neutral molecular species, to calculate the ambipolar diffusion
length scale and the strength of the plane-of-the-sky component of
the ambient magnetic field ($B_{\mathrm{pos}}$) in molecular clouds.
They studied the turbulent velocity dispersion spectra of a pair of
coexistent ion and neutral molecular species (HCN and $\mathrm{HCO^{+}}$)
in M17, with the assumption that the turbulent energy dissipation
process is associated with ambipolar diffusion. 

In this paper, we explain the method of \citet{li&houde} and our
purpose to test their turbulent energy dissipation model using optically
thin coexistent ion and neutral molecular species in $\S 2$, and
present our observations in $\S 3$. The data analysis is explained
in $\S 4$ and we end with a summary in $\S 5$.

\section{The Method of \citet{li&houde}}

\citet{ostriker} introduced a numerical method to study the kinetic
structure in a single-fluid turbulent cloud by measuring the distribution
of the velocity dispersion against the size and mass of the simulated
regions over which it was averaged. They showed how the lower envelope
of the calculated velocity dispersions depicts the actual turbulent
velocity spectrum of the gas in their simulations. \citet{li&houde}
used this technique to study the velocity dispersions of HCN ($J=4\rightarrow3$)
and $\mathrm{HCO^{+}}$ ($J=4\rightarrow3$) spectral line profiles
in M17 as a function of length scale. They noticed that the lower
envelope of the $\mathrm{HCO^{+}}$ data was consistently down-shifted
from that of HCN by a constant amount, independent of scale, and interpreted
this result as the signature of turbulent ambipolar diffusion acting
at smaller spatial scales than probed by their observations. Based
on these results \citet{li&houde} suggested a scenario where ambipolar
diffusion sets in at small enough length scales where neutrals decouple
from the flux-frozen ions and the magnetic field, and the friction
between the ions and the drifting neutrals causes the dissipation
of turbulent energy. \citet{tilley n balsara} tested this model by
simulating a two-fluid (ion and neutral) turbulence with a range of
different ionization fractions. Their results showed that at length
scales comparable to the ambipolar diffusion scale certain MHD waves,
including Alfvén waves, are strongly damped, and this damping causes
the velocity spectrum of ions to drop below that of the neutrals.
They simulated ion and neutral spectral line profiles for ionization
fractions ranging from $10^{-2}$ to $10^{-6}$ and plotted their
corresponding velocity dispersions against different beam sizes, producing
a result very similar to \citet{li&houde}. Thus they were successful
in numerically duplicating the observational results of \citet{li&houde},
including their determination of the turbulent ambipolar diffusion
scale.

Assuming the gas to be well coupled to the field lines at large scales
and that ion-neutral decoupling occurs at dissipation length scales,
\citet{li&houde} obtained an expression for $B_{\mathrm{pos}}$ by
setting the effective magnetic Reynolds number $R_{m}\sim V_{n}L/\beta$
to 1 at the decoupling length scale. Here $\beta=B^{2}/4\pi n_{i}\mu\nu_{i}$
is the effective magnetic diffusivity with $n_{i}$ the ion density,
$\nu_{i}$ the collision rate of an ion with the neutrals and $\mu$
the mean reduced mass characterizing such collisions. The aforementioned
expression for $B_{\mathrm{pos}}$ is

\begin{equation}
B_{\mathrm{pos}}=\left(\frac{L^{'}}{0.5\:\mathrm{mpc}}\right)^{1/2}\left(\frac{V_{n}^{'}}{1\:\mathrm{kms^{-1}}}\right)^{1/2}\left(\frac{n_{n}}{10^{6}\:\mathrm{cm^{-3}}}\right)\left(\frac{\chi_{e}}{10^{-7}}\right)^{1/2}\:\mathrm{mG},\label{eq:B}\end{equation}

\noindent where $L^{'}$ is the ion-neutral ambipolar diffusion decoupling
length scale, $V_{n}^{'}$ the neutral velocity dispersion at $L^{'}$,
$n_{n}$ the neutral volume density, and $\chi_{e}$ is the fractional
ionization. It should be noted that Equation (\ref{eq:B}) provides
only an approximate value for $B_{\mathrm{pos}}$ as it assumes that
ambipolar diffusion sets in exactly when $R_{m}=1$, which is most
likely too simple an approximation. The decoupling scale and the velocity
dispersion $V_{n}^{'}$ are determined by fitting the lower envelopes
of the square of the ion and neutral velocity dispersion data ($\sigma_{i}^{2}$
and $\sigma_{n}^{2}$, respectively) to Kolmogorov-type power laws.
If the coexisting ions and neutral species are well coupled at large
scales, their turbulent power spectra will have the same spectral
index over the inertial range (i.e., at scales larger than the dissipation
scale and smaller than the injection scale). Moreover, the cut-off
range for the ion and neutral energy dissipation do not coincide,
and assuming the ion energy spectrum to drop significantly at the
ambipolar diffusion scale, the difference between $\sigma_{i}^{2}$
and $\sigma_{n}^{2}$ is the difference of the integral of the ion
and neutral energy spectra between the ambipolar diffusion scale and
neutral viscosity dissipation scale (this difference, $a$ (see Equation
\ref{eq:sigma} below), is the shaded area in Figure 3 of \citealt{li&houde}).
With all the above assumptions, the $\sigma_{i}^{2}$ and $\sigma_{n}^{2}$
data are thus fitted to the following expressions

\begin{eqnarray}
\sigma_{i}^{2}(L) & = & bL^{n}+a\label{eq:sigma}\\
\sigma_{n}^{2}(L) & = & bL^{n}.\nonumber \end{eqnarray}

The observational determination of velocity dispersions from spectral
line profiles is a measurement of a velocity field that is integrated
along the line of sight within a column of gas of volume $l^{2}\times L$,
where $l$ is the beam size and $L$ is the depth through the cloud.
However, a given velocity field is associated with turbulent substructures
that span volumes different than $l^{2}\times L$. It should therefore
be examined whether the observational value of $\sigma^{2}$ at a
given beam size $l$ truly represents the actual velocity dispersion
at that length scale. \citet{Falceta} investigated this issue with
spectral line profile synthesis and three-dimensional MHD numerical
simulations. They showed that although the minimum observed velocity
dispersions scale as the actual $\sigma^{2}$, the coincidence of
individual values at a given scale depends on $l$ and the sonic Mach
number of the system. The observed dispersion is generally a good
approximation of the actual dispersion, although the latter is slightly
underestimated for supersonic turbulence and overestimated for a subsonic
regime. They also showed that the power law functions fitted to the
lower envelopes and actual three-dimensional velocity dispersions
from the simulations correspond well. The dispersions obtained using
the lower envelope therefore appear to be reliable estimates for the
actual velocity dispersions, as long as a large enough number of lines
of sight are considered.

\citet{li2010} studied the differences in the velocity dispersion
for the spectral lines of $\mathrm{H^{13}CN}$ ($J=1\rightarrow0$)
and $\mathrm{H^{13}CO^{+}}$ ($J=1\rightarrow0$) in DR21(OH) using
the data of \citet{lai_a}. They performed a similar analysis for
$\mathrm{HCN}$ ($J=3\rightarrow2$) and $\mathrm{HCO^{+}}$ ($J=3\rightarrow2$)
data they obtained with the Submillimeter Array in NGC 2024, and also
$\mathrm{HCN}$ ($J=4\rightarrow3$) and $\mathrm{HCO^{+}}$ ($J=4\rightarrow3$)
lines previously observed by \citet{Houde2002} in M17 and used in
\citet{li&houde}. In each source, the velocity dispersion comparisons
were performed for two locations with measured $B_{\mathrm{pos}}$
line densities. They found that the ion-neutral velocity difference
is almost constant in M17, which has a uniform density of magnetic
field lines, while in NGC 2024 and DR21(OH) the ion-neutral velocity
dispersion difference correlate well with the magnetic field line
density. This behavior is predicted and consistent with the model
of \citet{li&houde}. They also considered other possible parameters
that could cause differences between the velocity dispersion of the
two tracers, including optical depth, outflow enhancement, and presence
of hyperfine structures in the spectral lines, but only turbulent
ambipolar diffusion could consistently explain their observations. 

The spectra of an HCN and $\mathrm{HCO^{+}}$ pair, however, usually
appear to be optically thick and are subject to saturation. The aim
of this work is to test the turbulent energy dissipation model of
\citet{li&houde} with optically thin isotopologues i.e., H$^{13}$CN
and $\mathrm{H^{13}CO^{+}}$, to investigate whether the constant
down shift of the velocity dispersion spectrum of the ion from that
of the neutral is consistently reproducible with different species.
Accordingly, the lower optical depths of the aforementioned molecules
will allow us to potentially probe magnetic fields at greater depths
in star-forming regions, where they may be stronger \citep{li&houde,basu2000,crutcher99}.
Also, just like their main isotopologues they have very similar molecular
masses and the critical densities associated with their $J=4\rightarrow3$
transition are approximately $10^{7}$ $\mathrm{cm^{-3}}$ and $10^{6}$
$\mathrm{cm^{-3}}$, respectively. In particular, their line profiles
will not be saturated or self absorbed, which is likely to be the
case with HCN and $\mathrm{HCO^{+}}$, and therefore the calculated
line widths will be better representatives of the true velocity dispersions
characterizing the medium under study.

\section{Observations}

We obtained position-switch grid maps of $\mathrm{H^{13}CN}$$(J=4\rightarrow3)$
at 345.336 GHz and $\mathrm{H^{13}CO^{+}}$$(J=4\rightarrow3)$ at
346.998 GHz in weather grade 3 $(0.08<\tau_{225}<0.12)$ in DR21(OH)
during the months of August to November 2008 at the James Clerk Maxwell
Telescope (JCMT) on the summit of Mauna Kea, Hawaii. The observations
were performed using the HARP-B \citep{Buckle} single sideband receiver
and the ACSIS correlator configured for a bandwidth of 500 MHz and
channel resolution of $61$ kHz. HARP-B is a $4\times4$ element heterodyne
focal plane array that uses SIS detectors. The 16 detectors have receiver
temperatures of $94-165$ K and are separated by $30\arcsec$ with
a foot print of $2'$. To fully sample the $15\arcsec$ JCMT beam,
we mapped with sample offsets of $15\arcsec$ around the reference
position ($\alpha=20^{\mathrm{h}}39^{\mathrm{m}}01^{\mathrm{s}}$
and $\delta=42^{\circ}22^{'}37.7^{''}$, J2000) of the source. Since
the difference between the frequencies of $\mathrm{H^{13}CN}$$(J=4\rightarrow3)$
and $\mathrm{H^{13}CO^{+}}$$(J=4\rightarrow3)$ is 1.66 GHz, we made
use of the special confi{}guration capability of ACSIS to observe
the two spectra simultaneously within the 1.8 GHz passband of the
receiver. We used the previously published main beam efficiency of
$\simeq62$ \% determined with observations of Saturn at $345$ GHz
by \citet{Buckle}.

\section{Results}

\begin{figure}
\begin{center}  
\begin{tabular}{cc}  
\includegraphics[scale=0.6,angle=270]{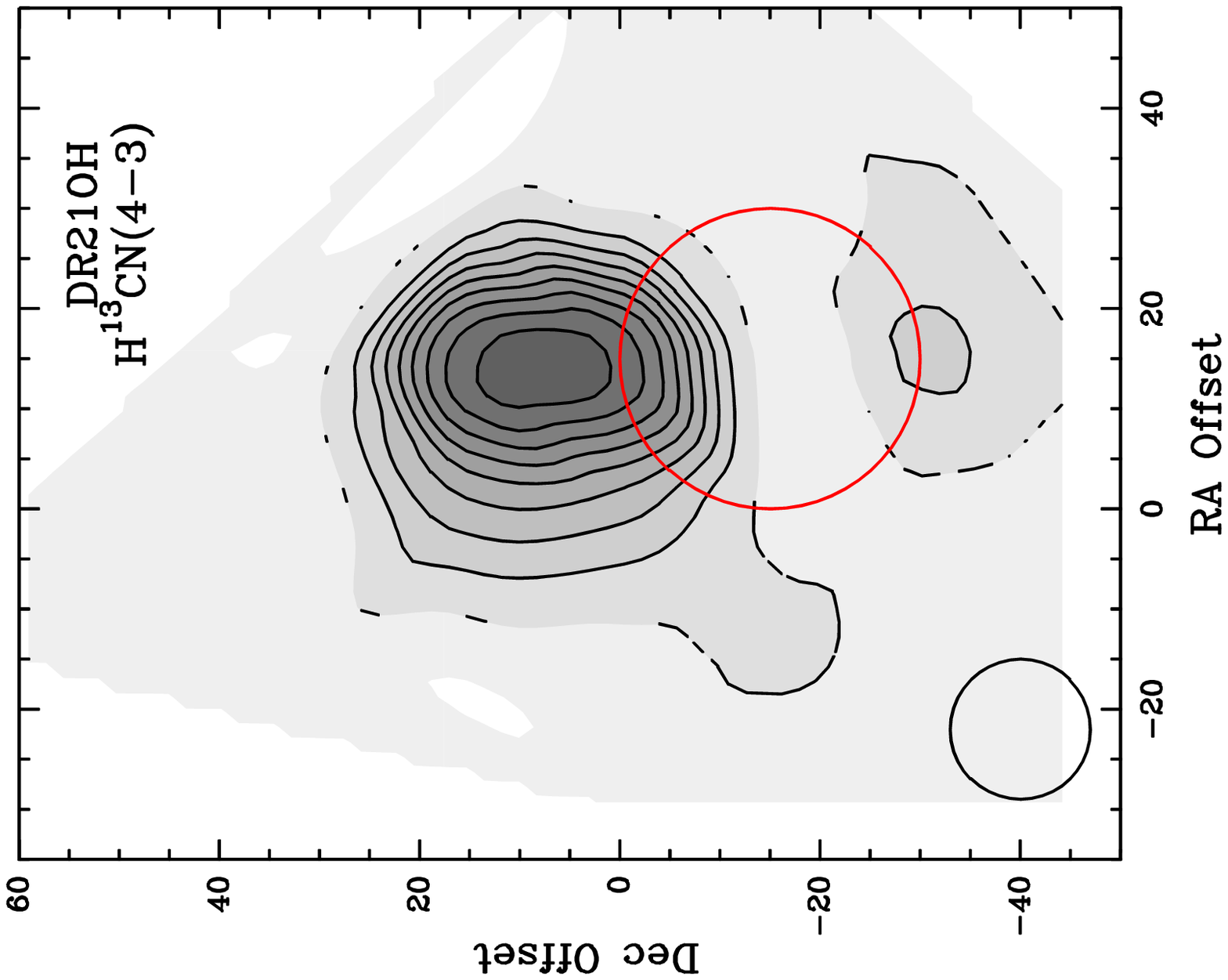} &              \includegraphics[scale=0.6,angle=270]{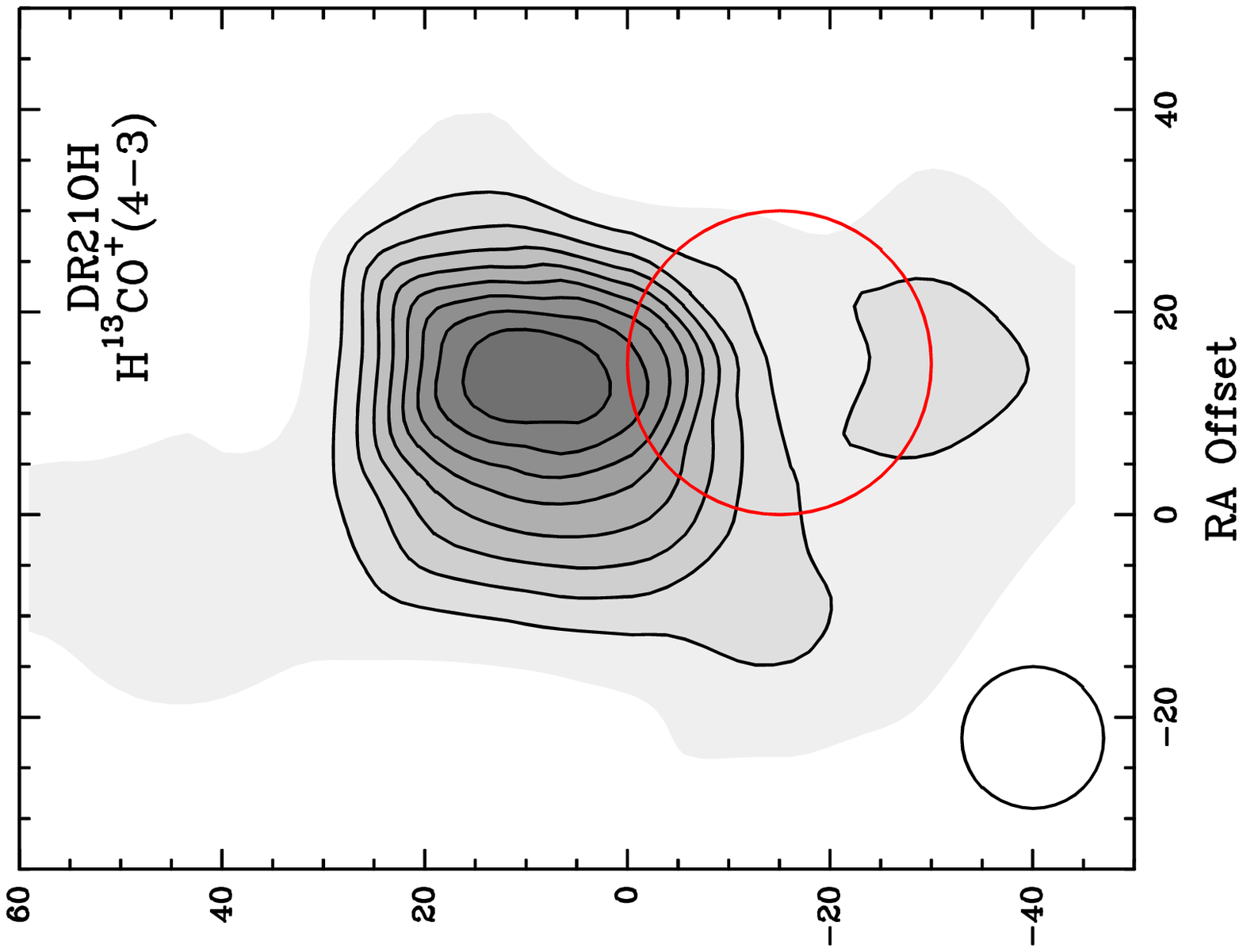} \\     
\end{tabular}   
\end{center}  
\caption{Comparison of the intensity maps of H$^{13}$CN and H$^{13}$CO$^{+}$   ($J=4\rightarrow3$) integrated over a velocity range of -20 $\mathrm{km\: s^{-1}}$  to 20 $\mathrm{km\: s^{-1}}$ in DR21(OH). The contours span a range of 10$\%$ to 90$\%$ of the peak (9.03 $\mathrm{K\:   km\: s^{-1}}$ for H$^{13}$CN and 8.23 $\mathrm{K\:  km\: s^{-1}}$ for H$^{13}$CO$^{+}$) by increments of 10$\%$. The red circles display the location along the line of sight where the minimum values of $\sigma^2$      are obtained observationally.} 
\label{fig:maps} 
\end{figure}

\begin{figure}
\epsscale{1.0}\plotone{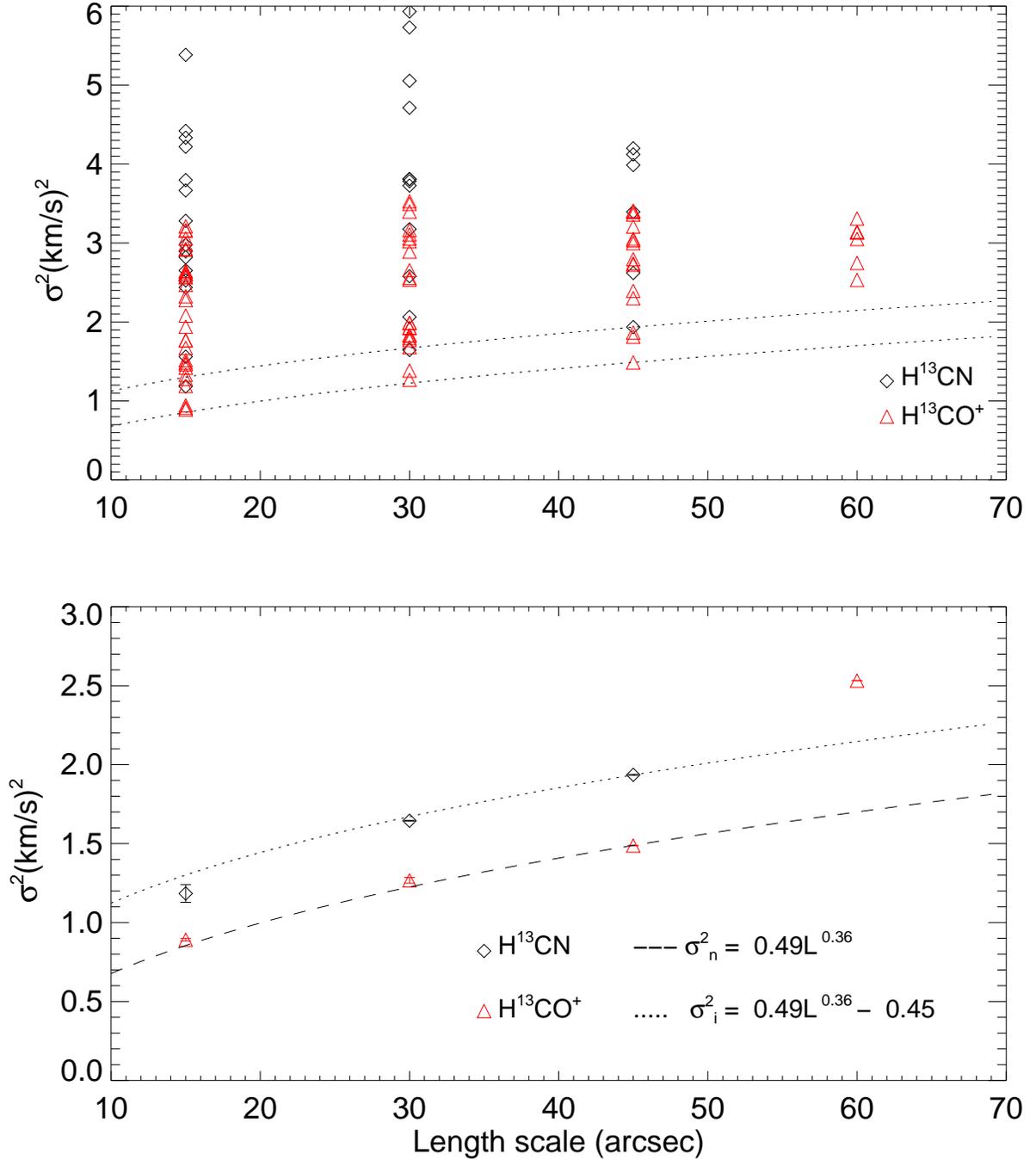}\caption{\label{fig:sigma2}$(Top)$ Plot of $\sigma^{2}$ as a function of
length scale in DR21(OH). The error bars for the data points are not
shown to avoid clutter in the graph. The values of $\sigma^{2}$ for
$\mathrm{H^{13}CN}$ at $60\arcsec$ fall above the plot margin. $(Bottom)$
The lower envelopes of the $\mathrm{H^{13}CN}$ and $\mathrm{H^{13}CO^{+}}$
data fitted for Kolmogorov-type power laws. }

\end{figure}

\begin{figure}
\begin{center}  
\begin{tabular}{cc}  
\includegraphics[scale=0.4]{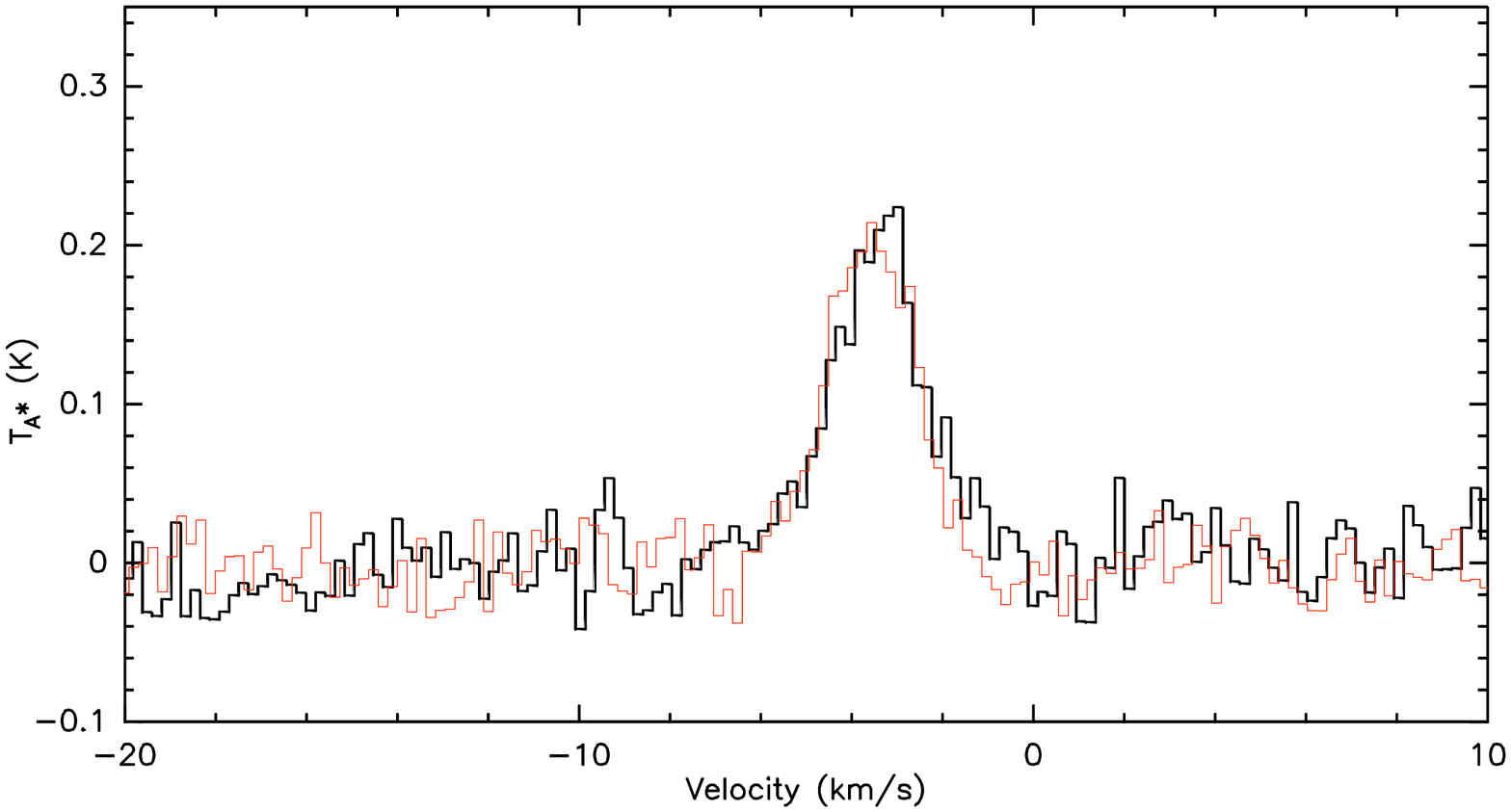} & \includegraphics[scale=0.4]{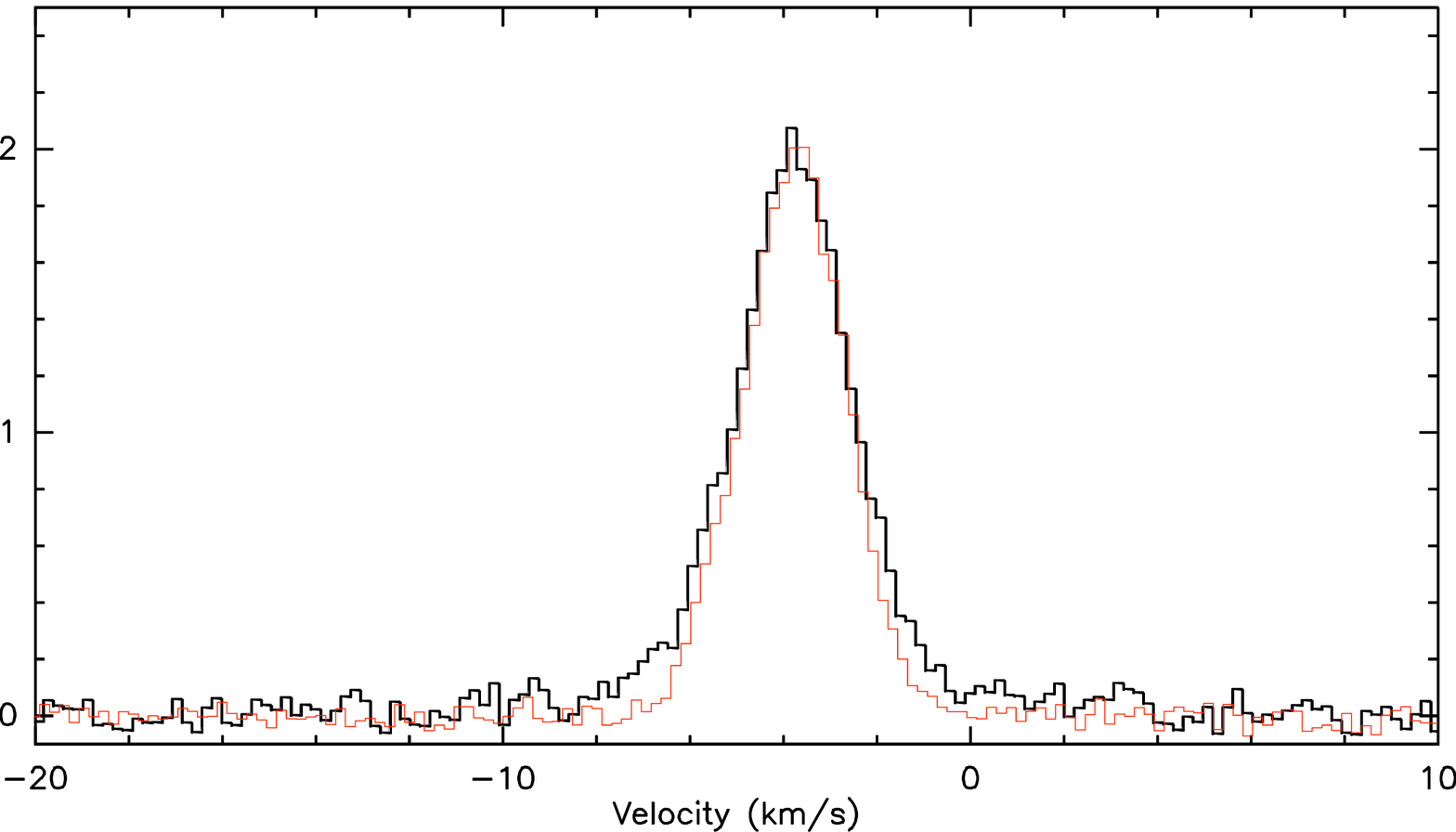} \\       \includegraphics[scale=0.4]{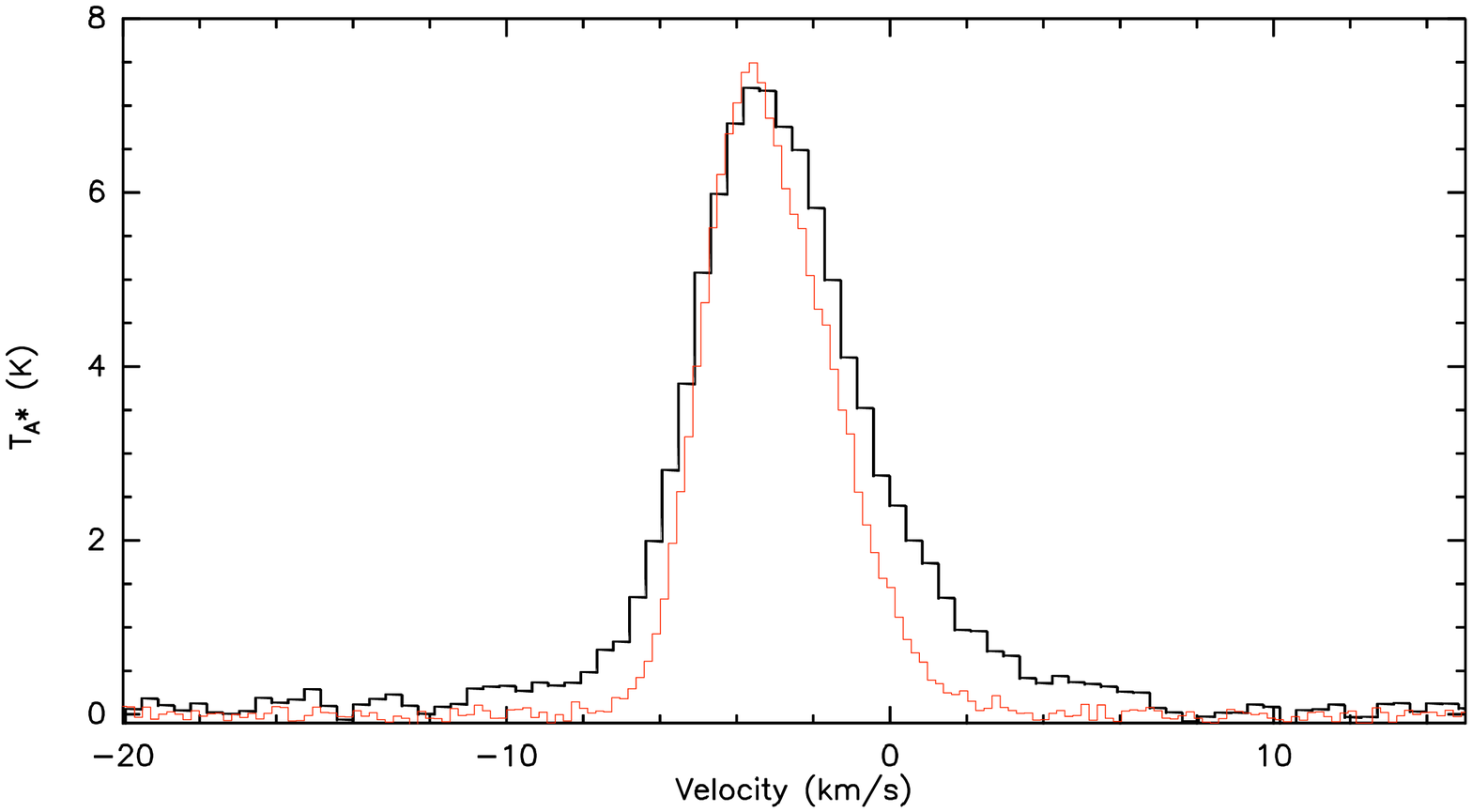} &       \includegraphics[scale=0.4]{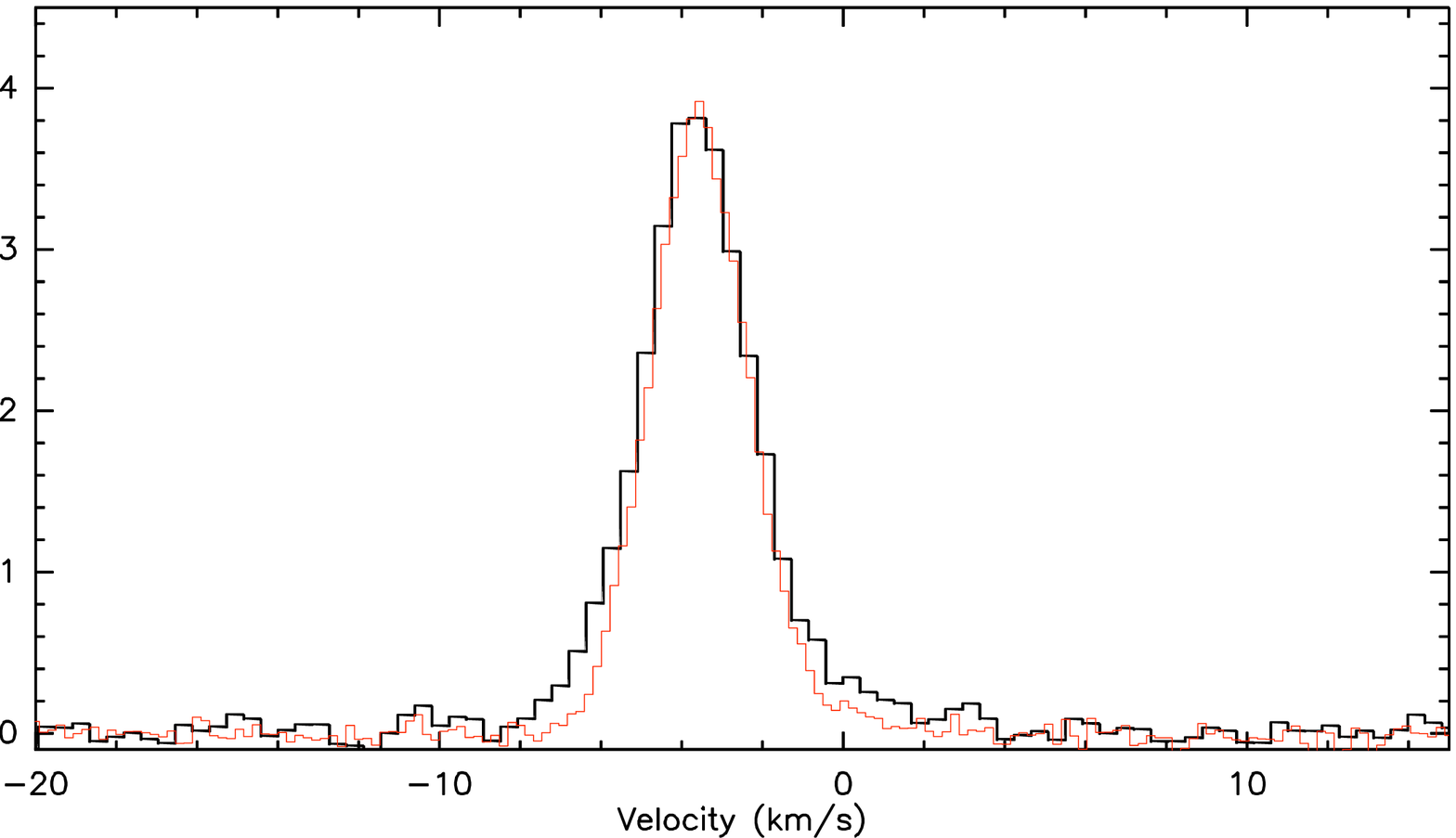} \\     
\end{tabular}   
\end{center}  
\caption{Comparison of the line widths of the $J=4\rightarrow3$ transition of H$^{13}$CO$^{+}$ (red) and H$^{13}$CN (black) spectral lines corresponding to the minimum $\sigma^2$ values. Starting from top left, going clockwise are the spectra at beam sizes 15", 30", 45" and 60". In all the figures, the H$^{13}$CN lines are scaled to the H$^{13}$CO$^{+}$ line temperatures to provide a  clearer comparison of the line widths.} 
\label{fig:spectra} 
\end{figure}

Figure \ref{fig:maps} shows the intensity maps of $\mathrm{H^{13}CN}$
($J=4\rightarrow3$) and $\mathrm{H^{13}CO^{+}}$ ($J=4\rightarrow3$),
integrated over a velocity range of $-20$ $\mathrm{km\: s^{-1}}$
to $20$ $\mathrm{km\: s^{-1}}$ in DR21(OH). The high critical density
of the $J=4\rightarrow3$ transition results in detection of radiation
from high density regions only, as can be seen from the intensity
contours congregating in an area of $\approx40\arcsec\times40\arcsec$
around the reference position of the source. We performed the spectral
line analysis with the Starlink%
\footnote{http://starlink.jach.hawaii.edu%
} and Gildas%
\footnote{http://iram.fr/IRAMFR/GILDAS/%
} software packages. The spectra of neighboring regions were averaged
across the maps to simulate beam sizes $15\arcsec,\:30\arcsec,\:45\arcsec$
and $60\arcsec$ at different lines of sight in the observed region.
We then fitted multi-Gaussian profiles to each spectrum and the corresponding
velocity dispersion $\sigma$ was then calculated using these fits.
The plot of the square of the velocity dispersions for which the dispersion
is at least three times larger than its corresponding uncertainty
is shown in the top panel of Figure \ref{fig:sigma2} as a function
of length scale, with the $\mathrm{H^{13}CN}$ data plotted in black
and $\mathrm{H^{13}CO^{+}}$ in red. We did not include the error
bars in this plot for a clearer representation of the data points.
The $\mathrm{H^{13}CN}$ data corresponding to the largest beam size
are not shown, as they do not fit in the limits of the plot; we had
low statistics at the $60\arcsec$ scale and therefore were not able
to produce quantitative data points for either of the species at this
scale. The presence of data points at each length scale above the
lower envelopes can be associated with the interception of the corresponding
line of sight with several turbulent substructures of different velocities
that brings an increase in the observed velocity dispersions. Therefore,
the minimum values for $\sigma^{2}$ may be attributed to the line
of sight that intercepts the fewest number of such turbulent sub-structures
\citep{Falceta}. All the minimum $\sigma^{2}$ values we obtained
for both species arise from one location in DR21(OH), circled in red
in Figure \ref{fig:maps}, thus further confirming the coexistence
of $\mathrm{H^{13}CN}$ and $\mathrm{H^{13}CO^{+}}$ and showing the
specific region along the line of sight that contains the least number
of turbulent substructures. The spectra for the different beam sizes
at this location are displayed in Figure \ref{fig:spectra}, with
the $\mathrm{H^{13}CO^{+}}$ profiles plotted in red and $\mathrm{H^{13}CN}$
in black and scaled to the temperature of $\mathrm{H^{13}CO^{+}}$
for a clearer comparison of their line widths. 

The lower envelopes of $\sigma^{2}$ for $\mathrm{H^{13}CN}$ and
$\mathrm{H^{13}CO^{+}}$ are plotted with their corresponding uncertainties
in the bottom panel of Figure \ref{fig:sigma2}, with Kolmogorov-type
power law fits given by Equation (\ref{eq:sigma}). As mentioned above,
we did not have sufficient statistics to find the true lower envelope
at the largest scale (i.e., $60\arcsec$) for both spectra and thus
used the $\sigma^{2}$ data at the three smallest beam sizes for the
power law fits. To enforce the same power law fit for both species,
we first fitted the difference between the square of the velocity
dispersion data of $\mathrm{H^{13}CN}$ and $\mathrm{H^{13}CO^{+}}$
to a constant function to obtain $a$, and then fitted the sum of
the two data sets to a power law of the form $2bL^{n}+a$ to obtain
$b$ and $n$. The fit results provide $a=-0.45\pm0.01$ $\mathrm{km^{2}\: s^{-2}}$,
$b=0.49\pm0.03$ $\mathrm{km^{2}\: s^{-2}}\:\mathrm{arcsecond^{-\mathit{n}}}$
and $n=0.36\pm0.02$. Although our fit is not as equally good for
the two species at all length scales as it was for \citet{li&houde},
it is still very satisfactory and perfectly adequate for our study.
Assuming a significant drop in the ion velocity spectrum at the dissipation
scale, the velocity dispersion of ions will be approximately equal
to $V_{n}^{'}$ at $L^{'}$ i.e., $V_{n}^{'}(L^{'})\simeq\sigma_{i}^{2}(L^{'})$.
We can then write \citep{li&houde}

\begin{eqnarray}
V_{n}^{'2}(L^{'}) & \simeq & 0.37bnL^{'n}\label{eq:v'}\\
 & \simeq & a+bL^{'}\nonumber \end{eqnarray}

and therefore

\begin{equation}
L^{'n}=-a/[b(1-0.37n)].\label{eq:L'}\end{equation}

Taking the distance from DR21(OH) to be 3 kpc \citep{genzel}, we
obtain $L'\simeq1.2\arcsec$ or $\simeq17$ mpc and $V_{n}^{'}\simeq0.26$
$\mathrm{km\: s^{-1}}$. 

We used the RADEX software package \citep{vandertak} to calculate
the gas volume density and optical depths associated with the $J=4\rightarrow3$
transition of $\mathrm{H^{13}CN}$ and $\mathrm{H^{13}CO^{+}}$. Taking
the values for the line widths and line temperatures of the spectral
lines observed at the reference position of the source, and the column
densities and excitation temperatures of the two species ($\simeq10^{13}$
$\mathrm{cm^{-2}}$ and $\simeq13$ K, respectively) from \citet{hezareh},
we get $n_{n}\simeq10^{6}$ $\mathrm{cm^{-3}}$ and assuming the observed
core to be as deep as it is wide, i.e., $40\arcsec$ , we obtain $N(\mathrm{H}_{2})\simeq2\times10^{24}$
$\mathrm{cm^{-3}}$. Additionally, RADEX computes optical depths of
0.17 and 0.21 for $\mathrm{H^{13}CN}$ and $\mathrm{H^{13}CO^{+}}$
lines, respectively. This optical depth comparison agrees with the
discussion by \citet{li2010} mentioned earlier. We also use $\chi_{e}=3.2\times10^{-8}$
for the ionization fraction as obtained by \citet{hezareh} for DR21(OH).
Applying the parameters above in Equation (\ref{eq:B}), we determine
a value of $B_{\mathrm{pos}}\simeq1.7$ mG. This value can be inaccurate
by a factor of a few due to the uncertainties in $n_{n}$ and $\chi_{e}$,
and the fact that Equation (\ref{eq:B}) is derived with the aforementioned
assumption of determining the turbulent ambipolar diffusion length
scale by setting $R_{m}=1$. 

Our results are in general consistent with the calculations of \citet{li&houde}
although the measurements were done on different sources and with
different pairs of molecular species. An exception is perhaps the
measured turbulent ambipolar dissipation scale for M17 reported by
\citet{li&houde} was $1.8$ mpc, almost an order of magnitude smaller
than what we measure for DR21(OH). This, however, may simply be the
result of a stronger magnetic field for our source. Indeed, using
the same fractional ionization of $3.2\times10^{-8}$ for M17 yields
$B_{\mathrm{pos}}\simeq0.57$ mG for that source, approximately three
times weaker than for DR21(OH). A relatively weaker field strength
for M17 than for DR21(OH) is also corroborated with existing CN $\left(N=1\rightarrow0\right)$
Zeeman measurements (\citealt{crutcher2,Falgarone}). Furthermore,
our result for $B_{\mathrm{pos}}$ is consistent with that of \citet{lai_b},
who obtained a polarization map of dust emission in DR21(OH) and used
the \citet{CH} method to calculate $B_{\mathrm{pos}}\simeq1$ mG.


We can now combine our value for $B_{\mathrm{pos}}$ with the Zeeman
measurement of $B_{\mathrm{los}}\simeq0.4$ mG by \citet{crutcher2}
in DR21(OH) and obtain a total strength of $B\simeq1.8$ mG at a volume
density $n(\mathrm{H}_{2})\simeq10^{6}\mathrm{\: cm^{-3}}$. We can
also use this result to calculate a mass to flux ratio for the observed
source. Taking $M/\Phi_{B}=1.0\times10^{-20}N(\mathrm{H}_{2})/|B|$
$\mathrm{cm^{2}\:\mu G}$ \citep{MS} with $N(\mathrm{H}_{2})\simeq2\times10^{24}$
$\mathrm{cm^{-2}}$ , we obtain $M/\Phi_{B}\simeq10$, which implies
that DR21(OH) exhibits a highly supercritical core. Although this
number could be overestimated by the column density considering our
assumption for the core geometry, it is consistent with the fact that
DR21(OH) has indeed revealed signs of high-mass star formation. Detection
of OH masers \citep{norris}, $\mathrm{H_{2}O}$ masers \citep{genzel},
and high-velocity outflows from its two dense cores, MM1 and MM2 \citep{lai_b},
are indications of ongoing star formation in this source.

\section{Summary \& Discussion}

We simultaneously observed and mapped $\mathrm{H^{13}CN}$ ($J=4\rightarrow3$)
and $\mathrm{H^{13}CO^{+}}$ ($J=4\rightarrow3$) in DR21(OH) to test
the turbulent energy dissipation model of \citet{li&houde}. The down
shift of the ion velocity dispersion spectrum from that of the neutral
is readily explained by the presence of ambipolar diffusion that causes
a steeper turbulent power spectrum for the ions compared to that of
the neutrals at dissipation length scales \citep{li&houde}. We used
our observational data to calculate a turbulent ambipolar diffusion
scale of $L^{'}\simeq17$ mpc and a plane of the sky magnetic field
strength of $B_{\mathrm{pos}}\simeq1.7$ mG for this source. 

Our analysis shows that the difference between the velocity dispersion
of the optically thin $\mathrm{H^{13}CN}$ and $\mathrm{H^{13}CO^{+}}$
spectral lines at different length scales exhibits a trend similar
to the results originally obtained by \citet{li&houde} with the optically
thick main isotopologues. It therefore follows that the ion line narrowing
effect first discussed by \citet{houde2000a} and the ensuing conclusions
of \citet{li&houde} cannot be due to a relative optical depth effect
between the ion and neutral species. This further corroborates prior
results from \citet{houde2000b}, which first investigated this possibility,
and more recently those of \citet{li2010}.

Although, as previously mentioned that optically thin spectral lines
are preferable for this analysis, it should be noted that $\mathrm{H^{13}CN}$
and $\mathrm{H^{13}CO^{+}}$ are less abundant than the main species
and are more difficult to map. Indeed, we observed the aforementioned
pair in three additional sources DR21(Main), NGC2264 and NGC2068 but
DR21(OH) was the only source in which we obtained enough detections
away from the core to perform our analysis. Moreover, observing HCN
and $\mathrm{HCO^{+}}$ in regions where they are optically thick
in the line core should not be too detrimental to the line width calculations,
as the wings of such spectral lines are still optically thin and suitable
for this study. More precisely, since the wings are more heavily weighted
in line width calculations, the (multi-) Gaussian profiles fitted
to such sections, away from the line peaks can be used for obtaining
the velocity dispersions in both species \citep{li2010}.

Finally, this technique has the advantage that it is not restricted
to nearby sources, since the data analysis mainly involves line width
calculations and does not require a spatial resolution high enough
to resolve the ambipolar diffusion scale.

\acknowledgements{The authors thank the referee for providing helpful comments, and
also S. Basu and W.Dapp for thoughtful discussions. M. H.'s research
is funded through the NSERC Discovery Grant, Canada Research Chair,
Canada Foundation for Innovation, Ontario Innovation Trust, and Western's
Academic Development Fund programs. }

\clearpage

\end{document}